\documentclass[12pt]{article}
\usepackage{graphicx}
\usepackage{epsfig}
\usepackage{color}

\usepackage{relsize}
\def\babar{\mbox{\slshape B\kern-0.1em{\smaller A}\kern-0.1em
    B\kern-0.1em{\smaller A\kern-0.2em R}}}

\def\uubar {\ensuremath{u\overline u}}

\def\ddbar {\ensuremath{d\overline d}}

\def\ssbar {\ensuremath{s\overline s}}

\def\ccbar {\ensuremath{c\overline c}}

\def\pim   {\ensuremath{\pi^-}}

\def\Kbar  {\ensuremath{\kern 0.2em\overline{\kern -0.2em K}}}

\def\Kp    {\ensuremath{K^+}}

\def\Kstarb   {\ensuremath{{\Kbar}\kern0.03em {\raise.18ex\hbox{$^*$}}}}

\def\Kzb   {\ensuremath{{\Kbar}\kern0.03em {\raise.1ex\hbox{$^0$}}}}
\def\KzKzb {\ensuremath{K^0 {\kern -0.16em \Kzb}}}
\def\Dz    {\ensuremath{D^0}}
\def\Dbar  {\ensuremath{\kern 0.2em\overline{\kern -0.2em D}}}
\def\Dzb   {\ensuremath{{\Dbar}\kern0.03em {\raise.3ex\hbox{$^0$}}}}
\def\DzDzb {\ensuremath{D^0 {\kern -0.16em \Dzb}}}

\def\Dstarb   {\ensuremath{{\Dbar}\kern0.03em {\raise.18ex\hbox{$^*$}}}}

\def\Bz    {\ensuremath{B^0}}
\def\B     {\ensuremath{B}}
\def\Bbar  {\ensuremath{\kern 0.18em\overline{\kern -0.18em B}}}
\def\Bzyb   {\ensuremath{{\Bbar}\kern0.03em {\raise.3ex\hbox{$^0$}}}}
\def\Bzb   {\ensuremath{{\Bbar}\kern0.03em {\raise.3ex\hbox{$^0$}}}}

\def\BB    {\ensuremath{B\Bbar}} 
\def\BzBzb {\ensuremath{B^0 {\kern -0.16em \Bzb}}}

\mathchardef\Upsilon="7107
\def\Y#1S{\ensuremath{\Upsilon{(#1S)}}}

\def\FourS {\Y4S}

\mathchardef\Delta="7101
\mathchardef\Xi="7104
\mathchardef\Lambda="7103
\mathchardef\Sigma="7106
\mathchardef\Omega="710A
\def\Deltabar   {\ensuremath{\kern 0.25em\overline{\kern -0.25em \Delta}}{}}
\def\Lbar {\ensuremath{\kern 0.2em\overline{\kern -0.2em\Lambda\kern 0.05em}\kern-0.05em}{}}
\def\Sigbar{\ensuremath{\kern 0.2em\overline{\kern -0.2em \Sigma}}{}}
\def\Xibar{\ensuremath{\kern 0.2em\overline{\kern -0.2em \Xi}}{}}
\def\Obar{\ensuremath{\kern 0.2em\overline{\kern -0.2em \Omega}}{}}
\def\Nbar{\ensuremath{\kern 0.2em\overline{\kern -0.2em N}}{}}

\def\mes        {\mbox{$m_{\rm ES}$}}

\def\ev   {\ensuremath{\rm \,e\kern -0.08em V}}
\def\kev  {\ensuremath{\rm \,ke\kern -0.08em V}} 
\def\mev  {\ensuremath{\rm \,Me\kern -0.08em V}} 
\def\gev  {\ensuremath{\rm \,Ge\kern -0.08em V}} 
\def\gevc {\ensuremath{{\rm \,Ge\kern -0.08em V\!/}c}} 
\def\tev  {\ensuremath{\rm \,Te\kern -0.08em V}}
\def\mevc {\ensuremath{{\rm \,Me\kern -0.08em V\!/}c}} 
\def\gevcc{\ensuremath{{\rm \,Ge\kern -0.08em V\!/}c^2}} 
\def\mevcc{\ensuremath{{\rm \,Me\kern -0.08em V\!/}c^2}}

\def\invfb   {\ensuremath{\mbox{\,fb}^{-1}}}
\def\mus  {\ensuremath{\rm \,\mus}}

\def\mus        {\ensuremath{\,\mu{\rm s}}}    
\renewcommand{\bar}[1]{\overline{#1}}

\def\gsim{{~\raise.15em\hbox{$>$}\kern-.85em
          \lower.35em\hbox{$\sim$}~}}
\def\lsim{{~\raise.15em\hbox{$<$}\kern-.85em
          \lower.35em\hbox{$\sim$}~}}

\def\to                 {\ensuremath{\rightarrow}}

\def\pep2{PEP-II}

\def\chic1{\ensuremath{\chi_{c1}}}
\def\chic2{\ensuremath{\chi_{c2}}}
\def\chic3{\ensuremath{\chi_{c3}}}

\def\jetset74   {\mbox{\tt Jetset \hspace{-0.5em}7.\hspace{-0.2em}4}}

\def\er #1 #2 { $#1 \pm #2$ }

\def\de        {\ensuremath {\Delta E^{*}}}

\def\mpl #1 #2 #3 {Mod.~Phys.~Lett.~{\bf#1},\ #2 (#3)}
\def\npb  #1 #2 #3 {Nucl.~Phys.~B~{\bf#1},\ #2 (#3)}
\def\plb  #1 #2 #3 {Phys.~Lett.~B~{\bf#1},\ #2 (#3)}
\def\pr   #1 #2 #3 {Phys.~Rep.~{\bf#1},\ #2 (#3)}
\def\prd  #1 #2 #3 {Phys.~Rev.~D~{\bf#1},\ #2 (#3)}
\def\prl  #1 #2 #3 {Phys.~Rev.~Lett.~{\bf#1},\ #2 (#3)}
\def\RMP  #1 #2 #3 {Rev.~Mod.~Phys.~{\bf#1},\ #2 (#3)}
\def\zpc  #1 #2 #3 {Z.~Phys.~C~{\bf#1},\ #2 (#3)}
\def\nim  #1 #2 #3 {Nucl.~Instrum.~Methods~{\bf#1},\ #2 (#3)}
\def\nima  #1 #2 #3 {Nucl.~Instrum.~Methods~A.{\bf#1},\ #2 (#3)}
\def\epjc #1 #2 #3 {Euro.~Phys.~Jour~{\bf#1},\ #2 (#3)}
\def\rmp #1 #2 #3 {Rev.~Mod.~Phys~{\bf#1},\ #2 (#3)}
\def\npbps #1 #2 #3 {Nucl.~Phys.~B.~proc.~suppl~{\bf#1},\ #2 (#3)}
\def\progtp #1 #2 #3 {Prog.~Theo.~Phys~{\bf#1},\ #2 (#3)}
\def\etal{{\it et al.}}

\def\eeul      {\ensuremath {3.3\times10^{-7}}}
\def\mmul      {\ensuremath {2.0\times10^{-7}}}
\def\emul      {\ensuremath {2.1\times10^{-7}}}

\newcommand{\xspace}    {\hspace{0.1cm}}

\newcommand{\GeV}       {\ensuremath{\mathrm{GeV}}}

\newcommand{\Dsl}{\ensuremath{D\ell\xspace}}

\def\etal {{\it et al.}}

\def\bll {\ensuremath {\Bz \to l^{+} l^{-}}}
\def\bee {\ensuremath {\Bz \to e^{+} e^{-}}}
\def\bmm {\ensuremath {\Bz \to \mu^{+} \mu^{-}}}

\def\bem {\ensuremath {\Bz \to e^{\pm} \mu^{\mp}}}

\def\bknn {\ensuremath {B^- \to K^-\nu \bar{\nu}}}
\def\bpdln {\ensuremath {B^+ \to {\bar{D}}^0 \ell^+ {\nu} (X)}}
\def\de  {\ensuremath {\Delta E}}
\def\mes {\ensuremath {m_{ES}}}

\def\beq{\begin{equation}}
\def\eeq#1{\label{#1}\end{equation}}
\def\eeqn{\end{equation}}

\def\beqa{\begin{eqnarray}}
\def\eeqa#1{\label{#1}\end{eqnarray}}
\def\eeqan{\end{eqnarray}}

\let\bar=\overbar

\def\etal{{\it et al.}}

\def\Dslash{\not{\hbox{\kern-4pt $D$}}}
\def\dslash{\not{\hbox{\kern-2pt $\del$}}}

\def\msb{{\bar{\ssstyle M \kern -1pt S}}}

\def\Title#1{\begin{center} {\Large {\bf #1} } \end{center}}

\begin{document}

\Title{New BaBar Results on Rare Leptonic B Decays}

\bigskip\bigskip


\begin{raggedright}  

{\it Valerie Halyo \index{Halyo, V.}\\
Stanford Linear Accelerator Center\\
2575 Sand Hill Rd.\\
Menlo Park, CA, 94025 \\
U.S.A}
\bigskip\bigskip
\end{raggedright}

\section{Abstract}

New preliminary BaBar results for rare leptonic decays 
$\bknn$ and $B^0\to \ell^+\ell^-$ are reported.
Using data collected  at the $\FourS$ with the 
BaBar detector, no  evidence for a signal was found yielding 
the corresponding  upper limits at the  $90\%$ confidence
level: ${\cal B}(B^-\to K^-\nu \bar{\nu})< 9.4\,\times\,10^{-5}$ for 
$50.7 \invfb\ $, ${\cal B}(\bee)< \eeul$, ${\cal B}(\bmm) < \mmul$ and 
${\cal B}(\bem) < \emul$ using $54.4  \invfb\ $.

\section{$\bknn$}
\begin{figure}[htb]
\begin{center}
\epsfig{file=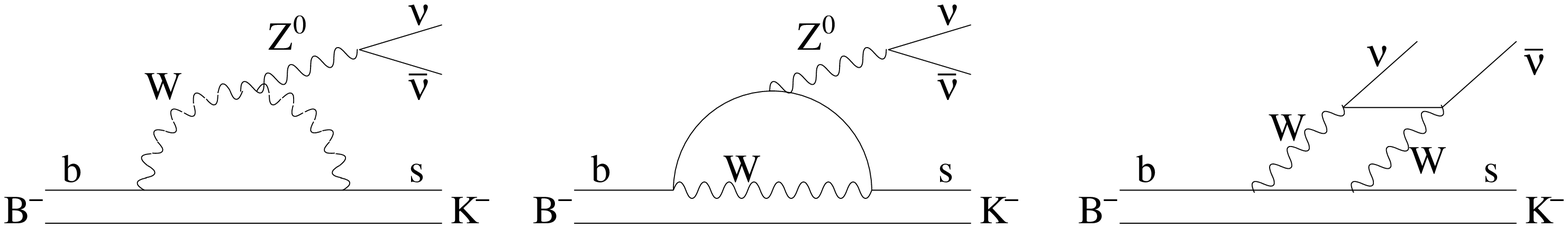,width=\textwidth}
\caption{Standard Model Feynman diagrams for $B^-\to K^-\nu \bar{\nu}$.}
\label{fig:bknn}
\end{center}
\end{figure}

The exclusive decay $\bknn$ is characterized by the absence of the long distance
contributions and by the fact that the effective Hamiltonian is represented 
in the Standard Model (SM) by only one operator. In the SM the decay proceeds
via the $W$ box and $Z$ penguin diagrams as can be seen in fig~\ref{fig:bknn}.
This mode probes the quark mixing parameter $|V_{ts}|$. The SM prediction
is ${\cal B}(B^-\to K^-\nu \bar{\nu}) = 3.8 ^{+1.2}_{-0.6} \times 10^{-6}$~\cite{btosnunu-theory2}\cite{smpre} and
the previous best limit obtained by CLEO is
 ${\cal B}(B^-\to K^-\nu \bar{\nu}) < 2.4 \times 10^{-4}$~\cite{cleo-bknunu}.
Enhancement beyond the SM can arise from various types of models~\cite{ligeti}.
The highly constrained models where the existing bounds on other 
Flavor Changing Neutral Current processes imply that
the rate for $\bknn$ can not exceed the SM prediction by more than a factor
of a few. In this category we have for example the Multi Higgs Doublet Models 
and  the Left Right symmetric models.
Enhancement by one or two orders of magnitude 
can arise from models with an extra vector-like down quark or models with
leptophobic $Z^{'}$ bosons~\cite{london}.
Last are the unconstrained models where the couplings 
responsible of enhancing $\bknn$ are to a large extent independent of
 current existing experiment bounds. An example which belongs to this category
is supersymmetric models without R-parity.

The presence of two neutrinos in the final state 
precludes the use of any kinematic constraints on the signal $B$ meson. 
The strategy adopted for the BaBar analysis was 
to look for a charged kaon with momentum $p^*>1.5\,\GeV$ in the ${\rm \FourS}$ frame recoiling against a semileptonic decay
{\bpdln} where X represent either a null or a photon or a $\pi^0$
from higher mass charm states. The exclusively reconstructed $B$ meson is 
referred in the following  as the tag $B$.
The BaBar standard practice of blind analysis was followed to prevent biases.
The low multiplicity of the signal decay reduces
the combinatorieal background in the tag $B$ reconstruction 
allowing the semileptonic decay {\bpdln} to be cleanly reconstructed.
The ${\bar{D}}^0$ is reconstructed in the $K^+ \pi^-$,  $K^+ \pi^- \pi^- \pi^+$
and $K^+ \pi^- \pi^0$ modes.  This method results in roughly 0.5\%\ of
tag reconstruction efficiency. 
The data used in the analysis consist of $50.7\invfb$ collected
at the ${\rm\FourS}$ resonance corresponding to $56.3\,\times\,10^6$
$\BB$ events and $6.4\invfb$ collected just below
$\BB$ threshold.
The kinematics for the $\bknn$ decays in the simulation is based on the form factor model 
in~\cite{btosnunu-theory2}.

Hadronic events were selected once an electron or muon 
with a momentum above $1.3\,\GeV$ in the {\rm \FourS} rest frame was identified.
Than ${\bar{D}}^0$ candidates were reconstructed in one of the 
$K^+ \pi^-$,  $K^+ \pi^- \pi^- \pi^+$
and $K^+ \pi^- \pi^0$ decay modes.
The kinematic requirement on the angle between the $B$ and the reconstructed $\Dsl$,
calculated in the ${\rm \FourS}$ frame was used to suppress background
and restrict the kinematics of the $ {\bar{D}}^0 \ell^+$
to be consistent with coming from a semileptonic $B$ decay. 
The requirment $-2.5<\cos\theta_{B~\Dsl}<1.1$ was imposed, using
\begin{equation}
\cos\theta_{B\,\Dsl} = \frac{2\,E_{{\B}} E_{{\Dsl}}
 -m^2_{{\B}} - m^2_{{\Dsl}}}
{2\,|\vec{p}_{{\B}}| |\vec{p}_{{\Dsl}}|   }\ \ .
\end{equation}
where $E_{\B}$ and $|\vec{p}_{\B}|$ are respectively 
the energy  and magnitude of 
the momentum of the $\B$ meson in the ${\rm \FourS}$ frame.

A special  double tag sample was used to extract a correction
to the efficiency calculated from signal Monte Carlo (MC).
The sample was reconstructed by first finding a
suitable ${\bar{D}}^0 \ell^+$ candidate where the ${\bar{D}}^0$ decays to $\Kp\pim$,
and then looking for a second $D\ell$ candidate in any of the
accepted ${\bar{D}}^0$ modes.
The observed rate of double tags per $\invfb$ in the data is $0.85\pm 0.11$ 
times the rate in the simulation leading to a correction to the 
the signal efficiency by a factor
$0.92\pm 0.06$ where the uncertainty is taken as a systematic error.
The uncertainty in the efficiency of several of the selection
criteria were also studied using the double-tagged sample. 
The total relative uncertainty on the selection efficiency was found to be
$\delta\epsilon/\epsilon = 8.7~\%$ where the tagging efficiency and 
($E_{\rm left}$) the remaing neutral energy after the
tag $B$ and its daughter were removed contribute the most.

\begin{figure}[htb]
\begin{center}
 \epsfig{file=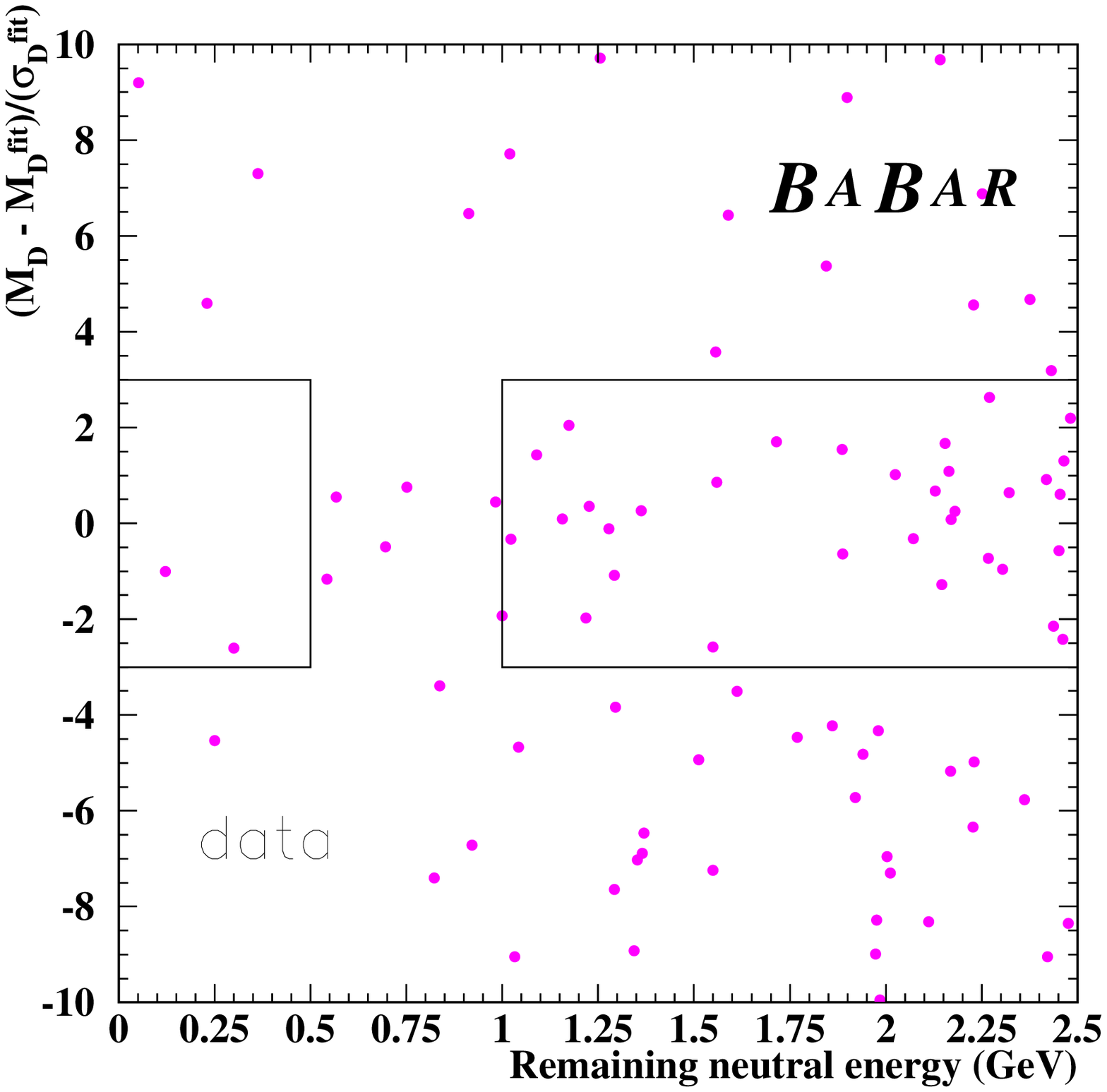,width=4.9cm,height=4.cm}
 \epsfig{file=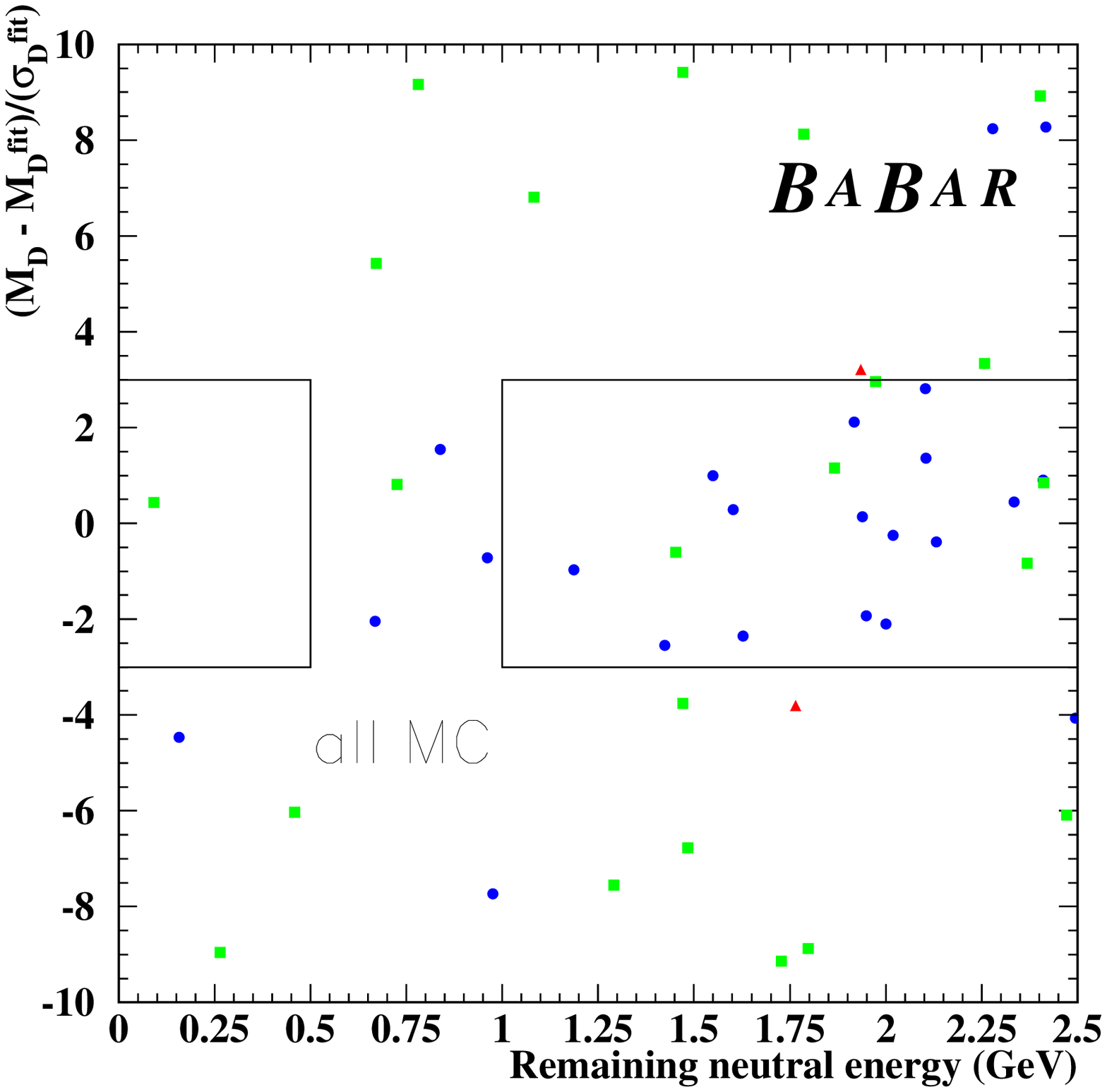,width=4.9cm,height=4.cm}
 \epsfig{file=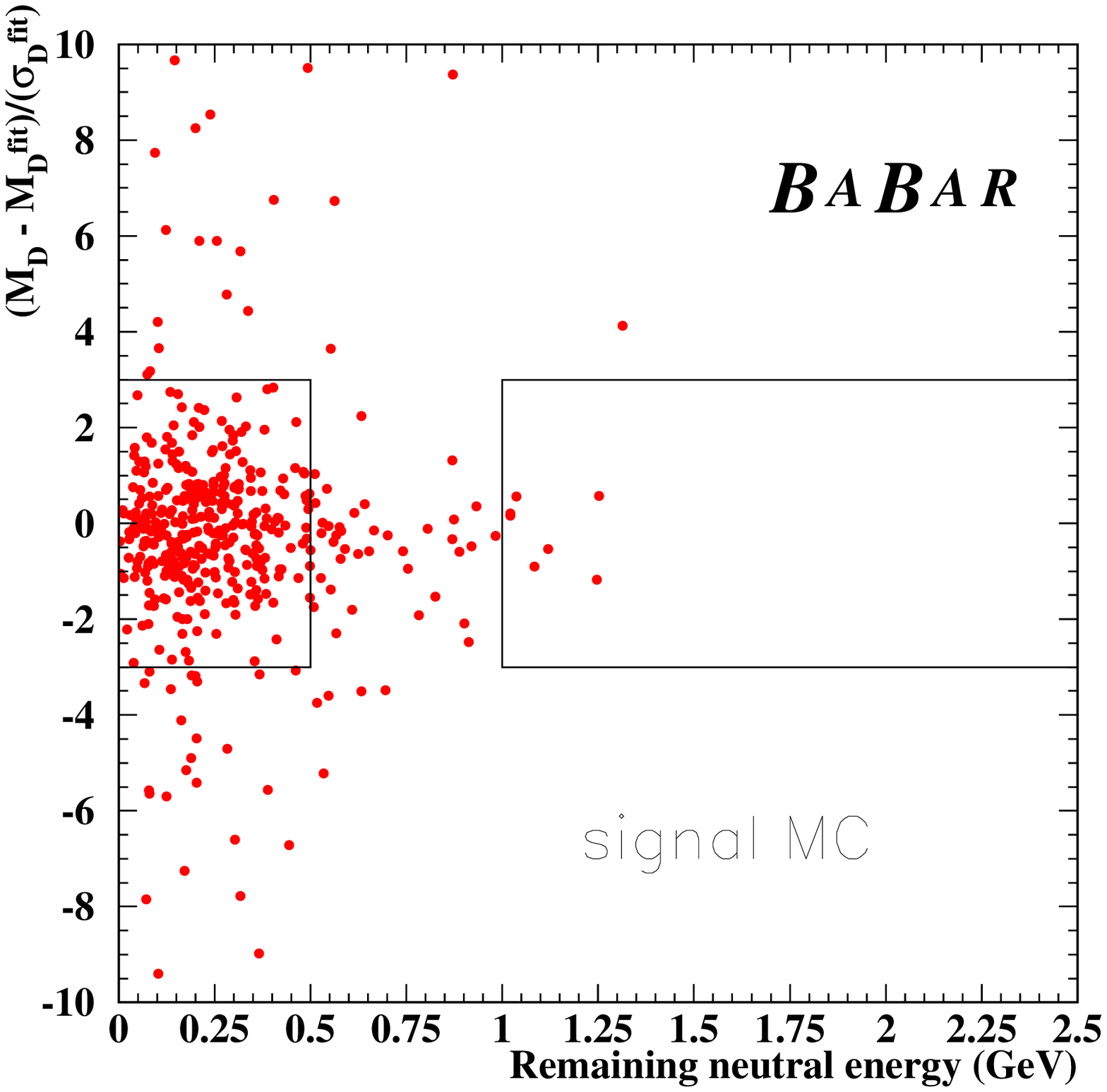,width=4.9cm,height=4.cm}
 \caption{The distribution of events in the 
 ($(m_D - m_D^{\rm fit})/\sigma_D^{\rm fit}$  $E_{\rm left}$)
 plane for on-peak data, generic $\BB$ and continuum MC and signal MC.
 In the generic MC plot the blue circles show the contribution
 from $\BB$ events, the green squares show the contribution from $\ccbar$
 and the red triangles show the contribution from $\uubar/\ddbar/\ssbar$.
 The generic MC contributions needs to be scaled by a factor of
 $1.09$, $2.21$ and $3.56$ for the $\BB$, $\ccbar$ and 
 $\uubar/\ddbar/\ssbar$ contributions respectively
 to correspond to the on-peak data luminosity.}
 \label{fig:boxs}
\end{center}
\end{figure}

The distribution of events in the search plane defined by the 
variables\footnote{The quantities $m_D^{\rm fit}$ and $\sigma_D^{\rm fit}$
are the mean and sigma from Gaussian fits to the $\Dz$ invariant mass
spectrum.  Separate values are calculated for each $\Dz$ decay mode in
data and simulation.$E_{\rm left}$ is the remaing neutral energy after the
tag $B$ and its daughter were removed }
$E_{\rm left}$ and $(m_D^{} - m_D^{\rm fit})/\sigma_D^{\rm fit}$ is
shown in fig~\ref{fig:boxs}.  We observe 2 events in the
signal box, defined by the requirements
$E_{\rm left}<0.5\,\GeV$ and $(m_D^{} - m_D^{\rm fit})<3\sigma_D^{\rm fit}$.
The expected background from the MC is 2.2 events.
The background at present appears to be mostly $ \ccbar$ events.

The Poisson upper limit calculated at $90\%$ C.L. without
 background subtraction  is
\begin{equation}
{\cal B}(B^-\to K^-\nu \bar{\nu}) < 9.4 \cdot 10^{-5}\,
\end{equation}

\vspace{-0.3in}
\section{$\bll$}
\begin{figure}[htb]
\begin{center}
\epsfig{file=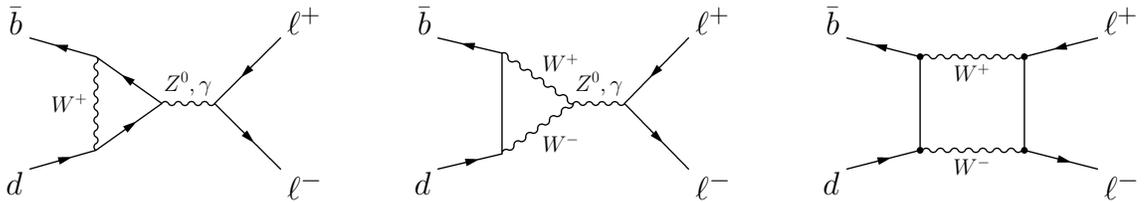,width=\textwidth}
\caption{Standard Model Feynman diagrams for $B^0\to \ell^+\ell^-$.}
\label{fig:bll}
\end{center}
\end{figure}

$\bll$ proceeds in the SM through  the three dominant 
$W$ box and $Z$ penguin diagrams shown in fig.~\ref{fig:bll}.
Even though these diagrams are similar to the one that lead to $\bknn$
these decays are further helicity suppressed  by factors of $m_{l}^{2}$.
Both $\bll$ and $\bknn$ FCNC transitions  provide an
essential  opportunity  to  test  the  SM and  offer  a  complementary
strategy in the search for new physics by probing the indirect effects
of new particles and interactions.
The SM theoretical   branching ratio  predictions  are   $1.9\times10^{-15}$ for
$\bee$, $8.0\times10^{-11}$ for $\bmm$ and null for $\bem$ although 
recent neutrino mixing experiment results suggest that the branching ratio would be less than $10^{-15}$.
To date these decays have  not been observed and the  current best limits from
CLEO and Belle are summarized in 
table~\ref{tab:results}~\cite{cleo-bll}-\cite{belle-bll}.  Since these
processes  are  highly  suppressed  in  the SM  they  are  potentially
sensitive probes of physics beyond the SM.
Models such as MHDM with Natural Flavor Conservation and large $tan \beta$
can give up to an order of magnitude enhancement~\cite{logan}. 
The enhancment in models with an extra vector-like down quark 
can be to up to two 
orders~\cite{Gronau}.
Large enhancement may also arise from Minimal Supersymmetric Models
with large  $tan \beta$~\cite{kolda} or Supersymmetric models without R-parity.

The decay  $\bll$ offers  a very clean  experimental signature  in the
BaBar  detector. The two high momentum leptons can be  measured 
precisely and identified with  high purity in
the  detector.   Relatively   low  backgrounds  arise  from  the continuum
 consisting mostly of non-resonant $e^+e^-
\rightarrow q \bar q$ production where $q=u,d,s,c$. The main
contribution  in   the  case   of  $\bee$  is   from  pairs   of  real
electrons in  $\ccbar$ production;  the  contribution of  misidentified
hadron-electron pairs  is negligible.  Two-photon  events contribute a
significant background.   Misidentification of muons is significantly
more important  in the  case of $\bmm$,  as evidenced by  an increased
background expectation  from $uds$ events. For the $\bmm$  channel, 
the background from two-photon processes is negligible.

The main measurement criteria are used to suppress different background events.
The multiplicity cut $N_{mult} = N_{Trk} + N_{\gamma}/2 \ge 6, E_\gamma > 80\mev$ 
suppresses  radiative Bhabha events while maintaining a higher 
efficiency than a simple stringent multiplicity cut.
The tracks are restricted to the central region of the detector using a polar angle cut 
which is efficient in removing QED background due to its strong dependence
 on the polar angle.
As mentioned above, the main background process comes from continuum events which 
exhibit a two jet structure and produce high momentum approximately back to back tracks 
satisfying the requirements imposed on our candidate events. 
Two different shape variables suppress this kind of background:
$|\cos\theta_T| < 0.84$ where $\theta_T$ is the angle between the thrust axes of the
$B$-candidate and the rest of the event and the event thrust magnitude $|T| < 0.9$.
The lepton pair  is selected by simultaneous
requirements on the energy difference $\de$ and the energy-substituted
mass $\mes$ defined in the following.
The invariant energy  difference of the $B$-meson  candidate and the 
 energy-substituted mass $m_{ES}$ are calculated as
 \linebreak
\begin{equation}
 \Delta E = {p_B \cdot p_i - s/2 \over \sqrt{s}} \;\;\;\;\; m_{ES} = \sqrt{(s/2 + {\bf p}_B\cdot{\bf p}_i)^2/E_i^2 - {\bf p}_B^2} \nonumber
\end{equation}

\noindent where $\sqrt{s}$ is  the CMS energy, $p_B$ and $ p_i$ denote
 the four-momenta  of the $B$-meson  candidate and the  initial state.

The signal box is defined in the ($\mes$ $\de$) plane and was obtained
for each of the $\bll$ mode separately using an upper limit optimization.
The size of the signal and Grand Sideband (GSB) was
chosen to be roughly $[+2,  -2]\sigma$  of  the  expected 
 resolution in  $\de$  and  $[+2,-2]\sigma$ for $\mes$.
  In  the cases of $\bee$  and $\bem$, the
signal box size  in $\de$ were relaxed to  roughly $[+2, -3]\sigma$ and
$[+2,  -2.5]\sigma$ to account  for increased  amounts of  final state
radiation  and  bremsstrahlung.

\begin{figure}[htb]
\begin{center}
 \epsfig{file=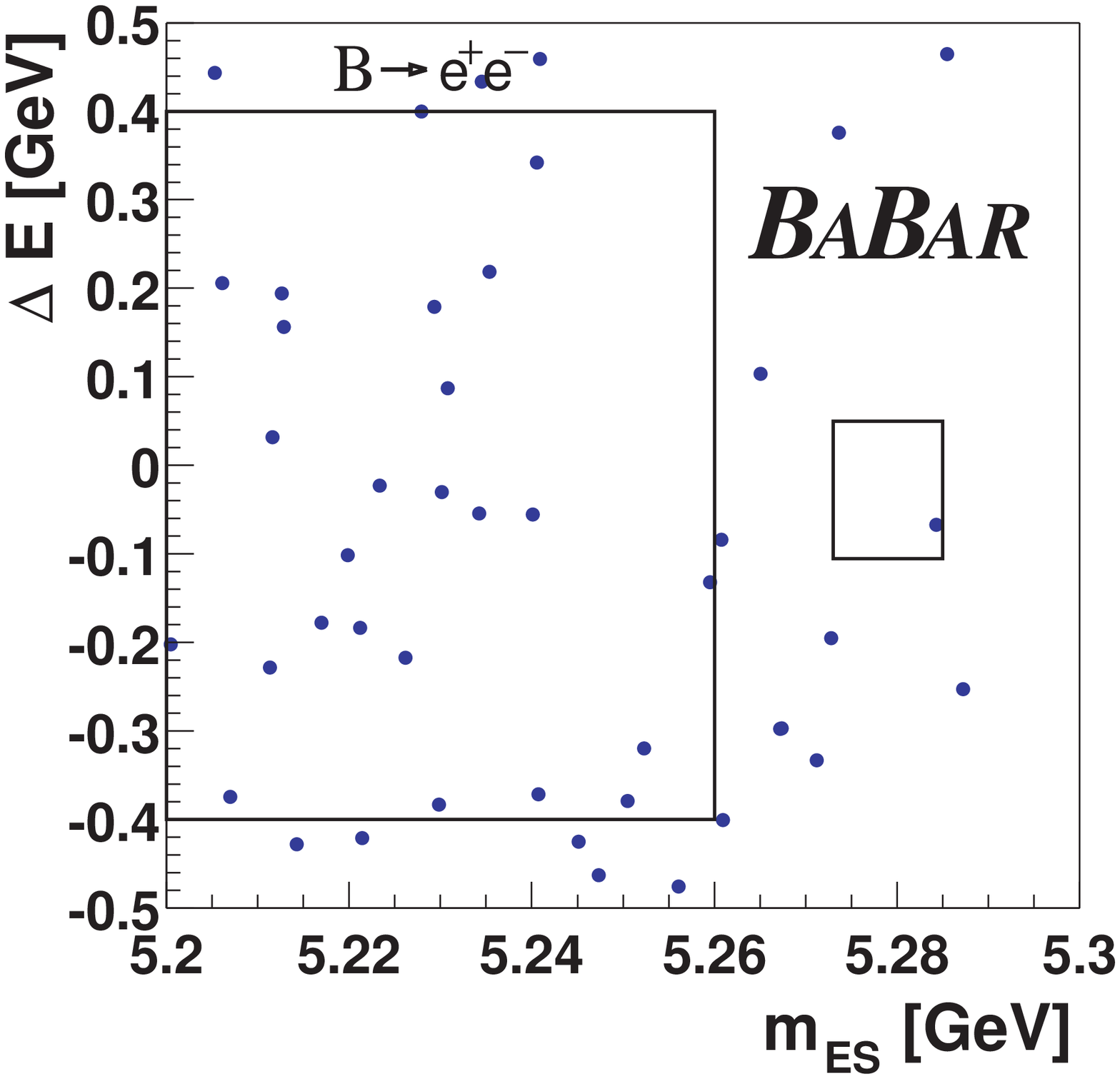,width=4.9cm,height=4.cm}
 \epsfig{file=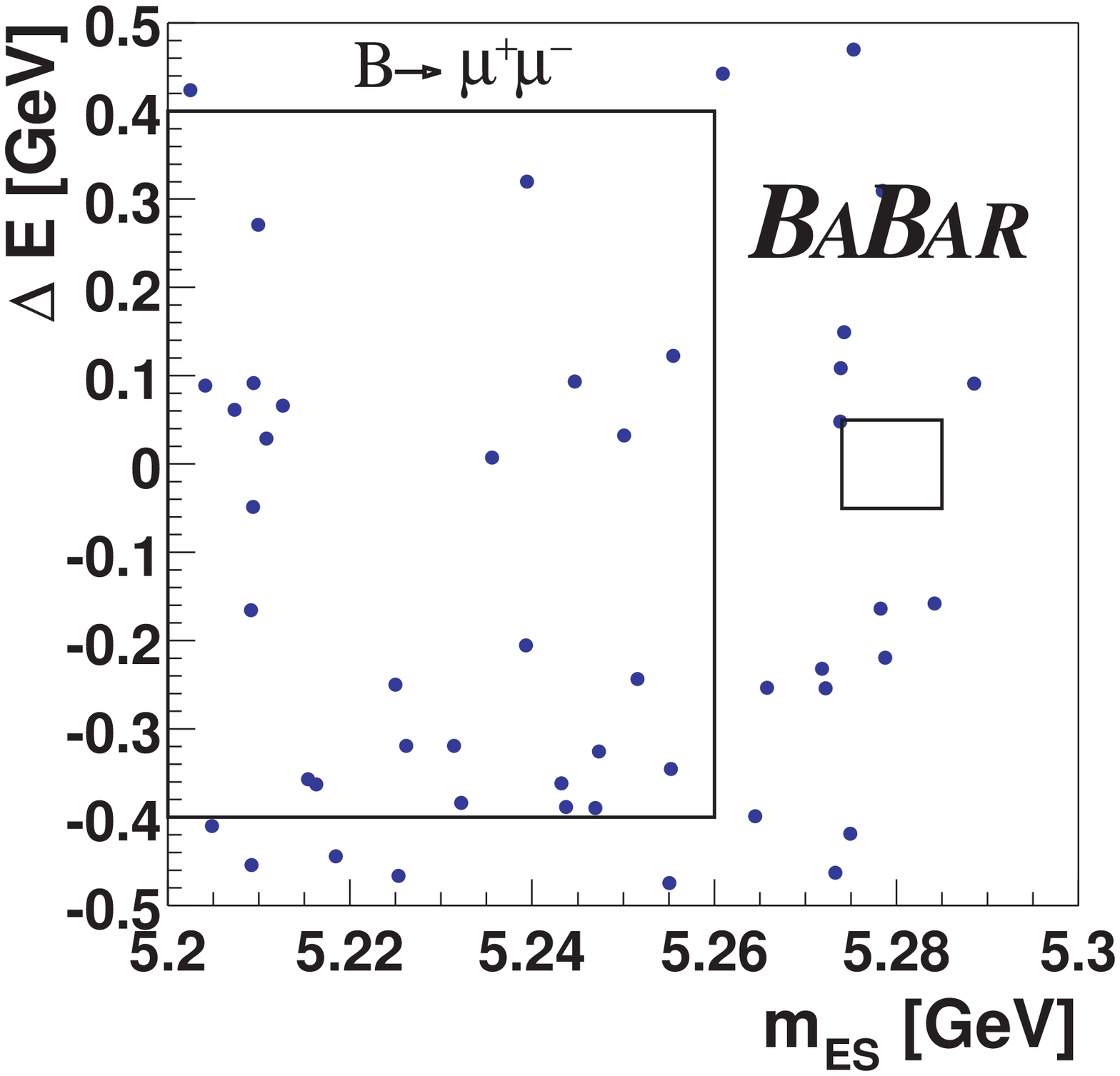,width=4.9cm,height=4.cm}
 \epsfig{file=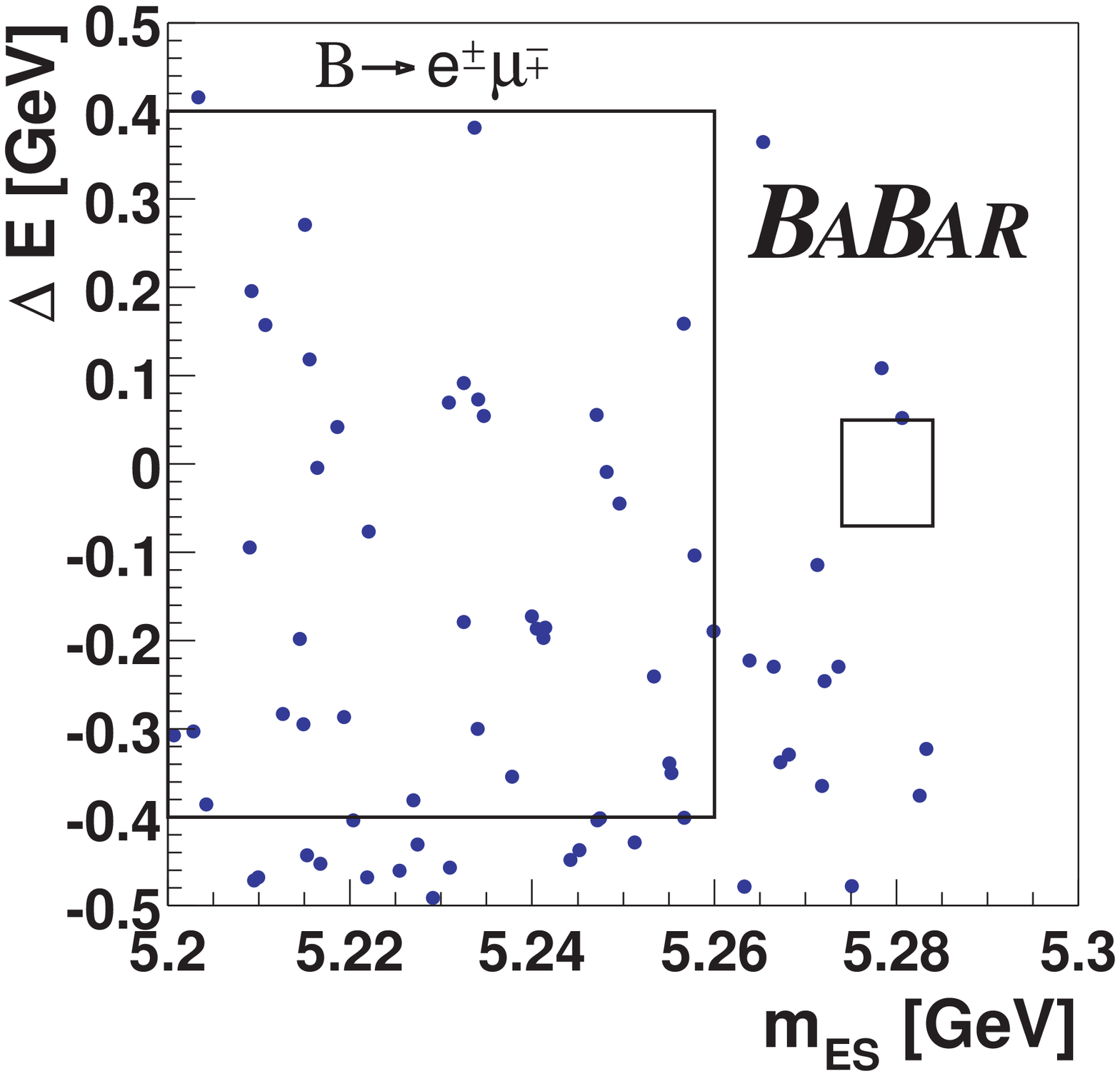,width=4.9cm,height=4.cm}
\caption{Unblinded $(\mes, \de)$ distributions with
 25, 26 and 37 events in the GSB for $\bee$,$\bmm$ and $\bem$ respectively.}
\label{fig:unblind}
\end{center}
\end{figure}

The resulting efficiencies for the three $\bll$ are given
 in table~\ref{tab:bllresult}.
The number of events observed in the GSB  appears in figure~\ref{fig:unblind}.
In order to estimate the number of background events in the signal box
an unbinned maximum likelihood fit was done using an 
Argus fit for the $\mes$ distribution and an exponential fit for $\de$
distribution. The results are given in table~\ref{tab:bllresult}.
There are three sources of systematic uncertainties: The normalization,
the signal efficiency and the background estimate. 
The systematic uncertainties for the different variables
were estimated using a control sample
of $B\rightarrow J/\Psi K^0_s$ with $J/\Psi \rightarrow \ell^+ \ell^-$.
Since there is no control sample for $\bem$ the error for $\bee$ 
was assigned to $\bem$
and the total systematic errors amount to $8.6\%$, $5.2\%$ and $8.6\%$ for
 $\bee$,$\bmm$ and $\bem$ respectively where the main contribution is from $\mes$ and $\de$.

\vspace{-0.2in}
After  unblinding,  we  observed  1   event  in  the  signal  box  
for $\bee$ and none for the rest of the channels are summarized in
table~\ref{tab:bllresult}.
The unblinded $(\mes, \de)$ distributions for the three channels
are shown in figs [\ref{fig:unblind}]. 
The observation are  compatible with  the  expected background.
We do not perform background  subtraction for the determination of the
branching fraction (upper limit). 

The upper limits on the branching ratios for $\bll$
 obtained at the  $90\%$  confidence  level are summarized in table~\ref{tab:bllresult}. 

\begin{table}[htb]
	\begin{center}
	 {\small
	\begin{tabular}{|c|r|r|r|r|r|} \hline
\hspace*{\fill}Channel\hspace*{\fill}&\hspace*{\fill}$N_{exp}$\hspace*{\fill} &\hspace*{\fill}$N_{obs}$\hspace*{\fill}   &\hspace*{\fill}$N_{BG}$\hspace*{\fill} &\hspace*{\fill}$\varepsilon [\%]$\hspace*{\fill} & UL ($90\%$ CL) \\ \hline

$\bee$&$1\times10^{-8}$ &$1$ &$0.60\pm0.24$ &$19.3\pm0.40_{stat}\pm1.60_{syst}$ &$3.3\times10^{-7}$\\

$\bmm$&$4\times10^{-3}$ &$0$ &$0.49\pm0.19$ &$18.8\pm0.28_{stat}\pm2.00_{syst}$&$2.0\times10^{-7}$ \\

$\bem$& \hspace*{\fill}---\hspace*{\fill} &$0$ &$0.51\pm0.17$ &$18.3\pm0.38_{stat}\pm1.50_{syst}$ &$2.1\times10^{-7}$\\ \hline
	\end{tabular}}
	\caption{Summary  of analysis.  $N_{exp}$  is the  number of  expected
	signal events  assuming a branching fraction  of $10^{-15}$. $N_{obs}$
	is the number  of observed events in the signal  box.  $N_{BG}$ is the
	expected number of background events in the signal box. }
	\label{tab:bllresult}	
       \end{center}
       \end{table}
\vspace{-0.4in}
\section{Results}
A summary of the BaBar preliminary results for the rare leptonic B decays are given in table~\ref{tab:results} in comparison with CLEO and Belle results.

\definecolor{darkblue}{rgb}{0.0,0.0,0.69}
\definecolor{darkred}{rgb}{0.82,0.0,0.0}
\begin{table}[htb]
	\begin{center}
	\begin{tabular}{|c|c|c|c|} \hline 
   {\color{darkblue}Mode}&{\color{darkblue}CLEO}&{\color{darkblue}Belle}&{\color{darkblue}Babar} \\ \hline
 ${\cal B}(B^-\to K^-\nu \bar{\nu})$ & $2.4\times 10^{-4}$ & - & {\color{darkred}$9.4 \times 10^{-5}$} \\ \hline 
 ${\cal B}(\bee)$&$8.3\times 10^{-7}$&$6.3\times 10^{-7}$& {\color{darkred}$3.3\times 10^{-7}$}\\
 ${\cal B}(\bmm)$ &$6.1\times 10^{-7}$&$2.8\times 10^{-7}$& {\color{darkred}$2.0\times 10^{-7}$}\\
 ${\cal B}(\bem)$&$15.0\times 10^{-7}$&$9.4\times 10^{-7}$& {\color{darkred}$2.1\times 10^{-7}$}\\ \hline
 Luminosity & $9.1\,\invfb$ & $21.3\,\invfb$ & $54.4\,\invfb$ \\ \hline
	\end{tabular}
          \caption{Summary of BaBar, CLEO and Belle results for rare leptonic B decay.}
 \label{tab:results}
	\end{center}
\end{table}

\end{document}